# The pattern speed of the bar in NGC 936

Michael R. Merrifield[1] and Konrad Kuijken[2,3]

[1] *Dept. of Physics, Southampton University, S017 1BJ*
[2] *Kapteyn Instituut, P.O. Box 800, 9700 AV, Groningen, Netherlands*
[3] *Visiting Scientist, Dept. of Theoretical Physics, University of the Basque Country, Lejona, Spain*

**ABSTRACT**

We have used the Tremaine–Weinberg method to measure the angular speed of rotation for the bar in the SB0 galaxy NGC 936. With this technique, the bar's pattern speed, $\Omega_p$, can be derived from the luminosity and stellar-kinematic information in long-slit spectral observations taken parallel to the major axis of the galaxy. The kinematic measurement required is the mean line-of-sight velocity of all stellar light entering the slit. This quantity can only be calculated reliably if any asymmetry in the shape of the broadening function of the spectral lines is also measured, and so we present a method which allows for such asymmetry. The technique also returns a true measure of the RMS uncertainty in the estimate. Application of the analysis to a set of long-slit spectra of NGC 936 returns four separate measures of $\Omega_p$ which are mutually consistent. Combining these data produces a best estimate for the bar pattern speed of $\Omega_p = 60 \pm 14 \, \mathrm{km\,s^{-1}\,kpc^{-1}}$ (assuming a distance of 16.6 Mpc). This result refines the only previous attempt to make this measurement, which yielded an estimate for $\Omega_p$ in NGC 936 of $104 \pm 37 \, \mathrm{km s^{-1}\,kpc^{-1}}$ (Kent 1987). The new measurement places the co-rotation radius just beyond the end of the bar, in agreement with theoretical calculations.

## 1 INTRODUCTION

Bar-like structures are found at the centres of approximately 30 per cent of disk galaxies (Sellwood & Wilkinson 1993). It is believed that these features rotate rapidly and maintain their shapes over many rotation periods. Hence, a defining parameter of any galactic bar is the angular velocity at which it rotates, its "pattern speed" $\Omega_p$. Orbit calculations show that a bar can only be built self-consistently if it lies entirely within its co-rotation radius – the point at which the pattern rotates at the same speed as a star on a circular orbit at that radius (e.g. Contopoulos 1980). Indeed, N-body simulations confirm that long-lived bar structures can form in disk galaxies, and that they extend only to near the co-rotation radii of their host galaxies (e.g. Combes & Sanders 1981, Sellwood 1981).

If we are to relate these theoretical ideas to observations of real galaxies and model their properties in any detail, then we must measure their pattern speeds. Unfortunately, this quantity has proved somewhat elusive to observation. In the case of simulations, the pattern speed can be measured directly by following the motions of the bar over time, but we do not have this luxury for a snapshot view of a real galaxy. Several methods have been suggested for estimating the pattern speed from the available data. For example, morphological features around the bar such as rings have been associated with the various resonant locations (such as the Lindblad and 4:1 resonances), and this information has been inverted to infer the fundamental parameters of the bar including $\Omega_p$ (see, for example, Combes & Gerin 1985, or Buta 1986). Where there are extensive observations of gas kinematics, it is possible to try to match the observations to hydrodynamic models with varying pattern speeds; the model that fits best then provides an estimate of $\Omega_p$ (e.g. Sanders & Tubbs 1980). Although these approaches seem to work quite well [see Sellwood & Wilkinson (1993) for an extensive review, or Sempere et al. (1994) for a recent application to NGC 4321], some more direct model-independent measure of the pattern speed is clearly highly desirable.

Tremaine & Weinberg (1984, henceforth TW) derived just such a measure by invoking the continuity equation to obtain $\Omega_p$ from the kinematic properties of a barred galaxy which are accessible to observation. If the galaxy lies at an inclination $i$, and if cartesian $x$ and $y$ axes are aligned parallel to its apparent major and minor axes such that the nucleus is located at $x = x_0$, then $\Omega_p$ can be calculated

from the relation

$$\begin{aligned}\Omega_p &= \frac{1}{\sin i}\frac{\int_{-\infty}^{\infty} l(x)[\overline{v}_{\mathrm{los}}(x) - v_0]\,dx}{\int_{-\infty}^{\infty} l(x)(x - x_0)\,dx}\\ &= \frac{1}{\sin i}\frac{\langle \overline{v}_{\mathrm{los}}\rangle - v_0}{\langle x\rangle - x_0},\end{aligned} \quad (1)$$

where the integrals are performed along any cut parallel to the apparent major axis, $\overline{v}_{los}(x)$ is the mean line-of-sight velocity of the stars at position $x$ on the slit, $v_0$ is the systemic velocity of the galaxy, and angled brackets denote the average value along the slit weighted by $l(x)$, the luminosity of the stars at each point. All of the quantities on the right-hand side of this equation can, at least in principle, be measured directly from observation: the integrated luminosity in a long-slit spectrum parallel to the major axis, weighted by position $x$ along the slit, allows us to calculate $\langle x \rangle$; and the Doppler shifts in the spectral lines in the same spectrum can be used to measure $\langle \overline{v}_{\mathrm{los}}\rangle$. The inclination $i$ can be measured from photometric observations if it is assumed that the galaxy becomes intrinsically circular at large radii.

In practice, it has proved difficult to apply this formula to observational data. TW showed that the method could not be applied reliably to H I observations, at least partly because H I gets converted into $H_2$ and stars in galaxies, and so the equation of continuity is not applicable. They suggested that optical observations of SB0 galaxies might prove more fruitful, as there is little gas or star formation in such systems to confuse the issue. Kent (1987) made the first attempt to follow up this suggestion by obtaining images and long-slit spectra of the SB0 galaxy NGC 936. This galaxy, shown in Fig. 1, is at an ideal inclination of $\sim 41°$ (Kormendy 1984), and its bar position angle lies $\sim 45°$ from the major axis, which optimizes the signal in the integrals in equation (1). The galaxy also appears to be nicely symmetric about its center, and its outer isophotes are elliptical with constant axis ratio, consistent with a bar embedded in an otherwise axisymmetric disk. Kent obtained an estimate for NGC 936's pattern speed of $\Omega_p = 104 \pm 37$ km s$^{-1}$ kpc$^{-1}$. This value places the co-rotation radius within the bar, contrary to the predictions of simple bar models, although the large uncertainty means that the measurement is still just consistent with a co-rotation radius beyond the end of the bar.

Kent's spectroscopic data were obtained using image tube observations recorded on photographic plates, and he expressed the desirability of obtaining CCD spectra of higher quality in order to reduce the errors. A further large gain in the signal-to-noise ratio of the spectral data can be obtained by noting that $\overline{v}_{\mathrm{los}}(x)$ only occurs inside the integral in equation (1). The long slit data can therefore be co-added along the slit to produce a single high signal-to-noise ratio one-dimensional spectrum, and the mean Doppler shift in the spectral lines then gives exactly the weighted value of $\langle \overline{v}_{\mathrm{los}} \rangle$ required in the TW formula. The difficulty in this approach is that the spectral line profiles in the co-added spectrum will be broadened in a significantly asymmetric manner, reflecting the distribution of $\overline{v}_{\mathrm{los}}(x)$ along the slit. The mean velocity cannot, therefore, be extracted by simple techniques such as cross-correlation (Tonry & Davies 1979) which assume that any broadening of the galaxy's spec-

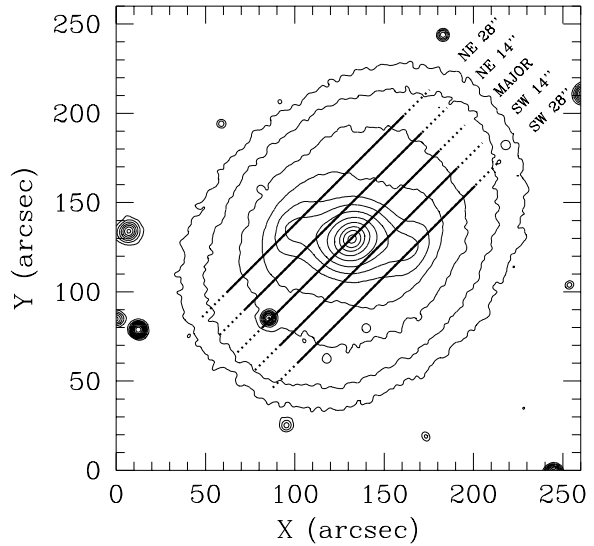

**Figure 1.** Contour plot of a I-band CCD image of NGC 936 obtained with the FLWO 48-inch telescope. The contours are spaced with 0.5 magnitude intervals. The slit positions for the spectral observations are overlayed: the solid line indicates the part of the slit that was co-added to produce the integrated mean velocity, and the dotted line shows the region that was used to define the background spectrum.

tral lines is symmetric. Recently developed techniques (e.g. Franx & Illingworth 1988, Bender 1990, van der Marel & Franx 1993, Kuijken & Merrifield 1993) permit us to obtain the complete shape of the Doppler broadening in a galaxy spectrum, and so we can now allow for asymmetric broadening to obtain the requisite unbiased estimate for $\langle \overline{v}_{\mathrm{los}} \rangle$.[⋆]

In this paper, we present the analysis of new CCD spectral observations of NGC 936 which allow us to calculate an improved estimate for its pattern speed. Section 2 describes the data, and the application of a stabilized spectral deconvolution method to obtain an unbiased estimate for the weighted mean value of $\langle \overline{v}_{\mathrm{los}} \rangle$. We also show how this method can be used to obtain a direct measure of the uncertainty in $\langle \overline{v}_{\mathrm{los}} \rangle$, and apply the technique to our data set. Combining this analysis with the derived value of $\langle x \rangle$ provides a well-constrained measure of the pattern speed in NGC 936, and we discuss the implications of the result in Section 3.

## 2 OBSERVATIONS AND ANALYSIS

We obtained long-slit spectra of NGC 936 on the night of 1993 December 9 using the Red Channel Spectrograph on

---

[⋆] As Kent (1987) recognised, this effect was also present when he analyzed the full two-dimensional spectrum, since it is unlikely that the stellar motions at any single point in a bar will result in a symmetric velocity distribution. The values for $\overline{v}_{\mathrm{los}}(x)$ derived by cross-correlation are therefore likely to be systematically incorrect. However, with the lower signal-to-noise ratio of the non-co-added spectra, it is impossible to correct reliably for any such asymmetries.

the Multiple Mirror Telescope at Mt. Hopkins, Arizona. The 12x8mmt CCD was used as a detector, and the spectrograph was configured with a 1200 lines per mm grating and a 1.25 arcsec × 3 arcmin slit, which provided a spectral resolution of 2.6Å (FWHM). The CCD pixels each correspond to 0.7Å, and so the instrumental broadening function is adequately over-sampled.

The spectral region near the Mg b triplet of lines ($\lambda \sim 5170$Å) was observed. The slit was aligned to be parallel to the major axis of the outer disk isophotes at a position angle of 135° (Kent 1987), and exposed at various offsets from the nucleus (see Table 1). The telescope was offset perpendicular to the slit position angle, ensuring that the zero point of position along the slit did not change between exposures: because of the near-perfect NW-SE alignment of the outer disk isophotes, these offsets were simply achieved by moving the telescope equal amounts N and E or S and W. To guard against loss of positional accuracy in these blind offsets, we confirmed that a final move back to the nucleus realigned the galaxy at its original position with respect to the slit.

Integration times were chosen so that the total signal obtained for each slit would be comparable: thus the outer, fainter positions were observed for somewhat longer (see Table 1). The five integrations were taken in sequence, with exposures of an arc lamp for wavelength calibration bracketing each one. Dome flats and biases were taken in the afternoons before and after the observations, and exposures of the twilight sky were taken in the early morning.

Standard data reduction was performed using IRAF. The spectra were corrected for bias and flat field variations. The bracketing arc lamp exposures were used to determine the wavelength scale of each galaxy spectrum, after which the data were rebinned on to a logarithmic wavelength scale running from 4760Å to 5660Å. The twilight sky exposures were used to determine the slit vignetting pattern on the detector, and this effect was then corrected for. In order to remove the night sky spectrum from the data, the spectra in the outermost 10 arcsec of slit were subtracted from the remainder of each frame. While there is still some flux from the galaxy at these radii, its contribution is quite minor; varying the exact region from which the sky background spectrum was extracted made no significant difference to any of the results.

The resulting reduced data frames were then analyzed as described below in order to yield the two averaged quantities required for each slit offset in the TW formula [equation (1)].

### 2.1 Mean positions along the slit

In order to calculate $\langle x \rangle$, the total flux in the slit between wavelengths of 4970Å and 5360Å was calculated as a function of position $x$ along the slit by simply summing the two-dimensional data in the spectral direction. The resulting slit luminosity profiles are shown in Figure 2. Mean positions were calculated from these profiles, and are marked on the figure and tabulated in Table 1. The statistical errors in $\langle x \rangle$ are totally negligible compared to those in $\langle \overline{v}_{los} \rangle$.

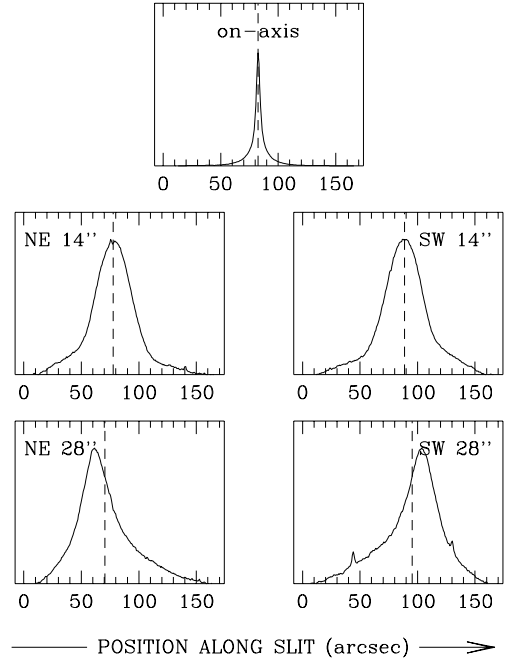

**Figure 2.** The total flux as a function of position along the slit, $x$, for the five integrations on NGC 936. The mean of $x$ in each case is indicated by a vertical dashed line.

### 2.2 Mean line-of-sight velocities

The second quantity required for the TW analysis, the mean luminosity-weighted line-of-sight velocity of the stars that illuminate the slit $\langle \overline{v}_{los} \rangle$, is rather less straightforward to calculate. For each slit position, we need to extract the Doppler broadening function of the spectra after they have been co-added in the spatial direction; the value of $\langle \overline{v}_{los} \rangle$ is then just the mean value of this distribution. As a first step, we therefore summed the spectra in each slit which lay less than 70 arcsec from the galaxy's minor axis.

We measured the broadening function of each of these co-added spectra using the Unresolved Gaussian Decomposition (UGD) algorithm (Kuijken & Merrifield 1993). Briefly, this technique involves a simple least squares fit to the continuum-subtracted galaxy spectrum $G(\ln \lambda)$. The model that we fit consists of a sum of Doppler-shifted copies of a single component spectrum $\overline{S}$:

$$G_{\rm mod}(\ln \lambda) = \sum_{j=1}^{N} a_j \overline{S}(\ln \lambda - v_j/c). \qquad (2)$$

The parameterization in terms of logarithmic wavelength turns the Doppler shift into a constant offset, and velocity broadening into a linear convolution. The spectrum $\overline{S}$ is formed by continuum-subtracting a stellar template spectrum and then broadening it by convolving it with a Gaussian of dispersion $\Delta$. The coefficients $a_j$ are optimized to give the closest match to the observed galaxy spectrum. The best-fit model is then the convolution of the original stellar template spectrum with the broadening function,

Table 1. Observations, analysis and results summary for the five long-slit spectra of NGC 936.

| Offset | exp. time (s) | $\langle x \rangle$ (arcsec) | $v_j$ (km s$^{-1}$) | $\Delta$ (km s$^{-1}$) | $\langle \overline{v}_{\rm los} \rangle$ (km s$^{-1}$) |
|---|---|---|---|---|---|
| 28″ NE | 1800 | 70.5 | 1155 (133.3) 1688 | 89 | 1377 ± 13 |
| 14″ NE | 1200 | 77.8 | 1066 (177.7) 1777 | 111 | 1400 ± 13 |
| nucleus | 600 | 82.6 | 977 (222.1) 1866 | 125 | 1425 ±  7 |
| 14″ SW | 1200 | 88.7 | 1066 (177.7) 1777 | 111 | 1434 ± 16 |
| 28″ SW | 1800 | 95.3 | 1155 (133.3) 1688 | 89 | 1452 ± 16 |

The entries for $v_j$ list the minimum, stepsize, and maximum velocities for the means of the unresolved gaussian components.

$$f(v_{los}) = \sum_{j=1}^{N} \frac{a_j}{\sqrt{2\pi}\Delta} \exp\left[-\frac{(v_{los} - v_j)^2}{2\Delta^2}\right]. \qquad (3)$$

By choosing the separation of the gaussians' means, $v_j$, to be slightly less than $2\Delta$, we can model galaxy spectra which have been doppler broadened by any function that is smooth on the scale of $\Delta$.

As well as smoothness, we place two further constraints on the models. Firstly, the broadening function must be non-negative at all velocities. This requirement arises from the physical requirement that the stars in the galaxy all make positive contributions to the total luminosity; it effectively suppresses high frequency noise in the derived broadening function. This linear constraint on the solution converts the least-squares analysis into a quadratic programming problem, which can be solved efficiently even for large numbers of components using a variation of the simplex algorithm. The algorithm also returns a matrix with components $C_{ij}$, the covariance between the fitted coefficients $a_i$ and $a_j$.

The second imposed constraint is that we use gaussians in only a limited range around the systemic velocity of the galaxy. We thus avoid spurious high-velocity features in the broadening function which might arise from crosstalk between different spectral lines. In order to choose the appropriate range for the broadening function to be non-zero, we have adopted a variant of the $3\sigma$ clipping algorithm which has previously been applied to velocity distributions of discrete objects such as galaxies in clusters (e.g. Yahil & Vidal 1977). We first fit a single gaussian broadening function to the data to obtain a crude estimate for the dispersion of the distribution, $\sigma_{est}$. We then carry out the full velocity distribution analysis using five gaussian components. We place the central component at the systemic velocity of the galaxy, $v_0$, and position the other four components uniformly on either side so as to cover the range between $v_0 - 3\sigma_{est}$ and $v_0 + 3\sigma_{est}$. The widths of the gaussians are set to approximately 0.6 times the separation between components. The adopted parameters are given in Table 1. Note that the symmetric choice of values for $v_j$ about the systemic velocity of the galaxy avoids biasing the derived means systematically: the outer parts of the galaxy, which are axisymmetric, should have velocity distributions which are symmetric about the systemic velocity, and our choice of components allows these outer parts of the slit to contribute equally to the broadening function.

The resulting broadening functions are shown in Fig. 3. Clearly, the positivity constraint and $3\sigma$ clipping have resulted in distributions from which robust estimates of the mean can be obtained. The functions are also obviously non-gaussian, and slits located on opposite sides of the galaxy show oppositely skewed profiles as expected in an equilibrium system.

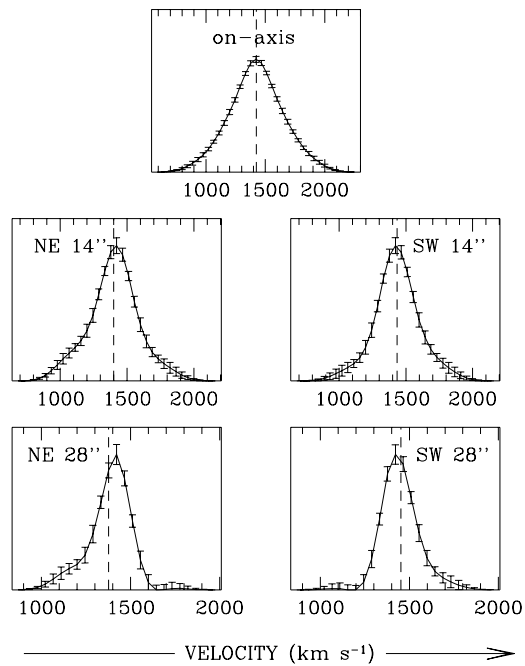

Figure 3. Line profiles for the coadded spectra for each offset position. The error bars, while correlated between adjacent pixels, are the 1-$\sigma$ error on the best-fit velocity distribution with the smoothing listed in Table 1. The mean of each distribution is shown with a vertical dashed line.

Given the best-fit amplitudes, $a_i$, for the gaussian components and their covariance matrix elements, $C_{ij}$, we can calculate the mean velocity of the broadening function and its error. The mean velocity is simply

$$\langle \overline{v}_{\rm los} \rangle = \sum_i a_i v_i \Big/ \sum_i a_i. \qquad (4)$$

Its error can be calculated by noting that the covariance between two linear combinations of the $a_i$'s is given by

$$\mathrm{Cov}\left(\sum_i p_i a_i, \sum_j q_j a_j\right) = \sum_{ij} p_i q_j C_{ij}. \qquad (5)$$

Using the result, valid for small variances, that $\mathrm{Var}(\ln X) = X^{-2}\mathrm{Var}(X)$, and setting $T = \sum_i a_i v_i$, $B = \sum_i a_i$, we find that

$$\begin{aligned}
&\mathrm{Var}(\langle \overline{v}_{\mathrm{los}}\rangle) \\
&= \langle \overline{v}_{\mathrm{los}}\rangle^2 \mathrm{Var}(\ln\langle \overline{v}_{\mathrm{los}}\rangle) \\
&= \langle \overline{v}_{\mathrm{los}}\rangle^2 \mathrm{Var}(\ln T - \ln B) \\
&= \langle \overline{v}_{\mathrm{los}}\rangle^2 \left[\mathrm{Var}(\ln T) - 2\mathrm{Cov}(\ln T, \ln B) + \mathrm{Var}(\ln B)\right] \\
&= B^{-2}\left[\mathrm{Var}(T) - 2\langle \overline{v}_{\mathrm{los}}\rangle\mathrm{Cov}(T,B) + \langle \overline{v}_{\mathrm{los}}\rangle^2 \mathrm{Var}(B)\right].
\end{aligned} \quad (6)$$

By adjusting the velocity zeropoint so that the mean velocity is zero, (an operation which does not affect the variance of $\langle \overline{v}_{\mathrm{los}}\rangle$), the last two terms can be made to vanish, leaving us with the equivalent but more compact expression,

$$\mathrm{Var}(\langle \overline{v}_{\mathrm{los}}\rangle) = \frac{\sum_{ij}(v_i - \langle \overline{v}_{\mathrm{los}}\rangle)(v_j - \langle \overline{v}_{\mathrm{los}}\rangle) C_{ij}}{\left(\sum_i a_i\right)^2}. \quad (7)$$

We can thus propagate the errors on the fit to the galaxy spectrum through the entire analysis to derive the uncertainty in the mean line-of-sight velocity.

We have applied this analysis to the five co-added spectra of NGC 936. The resulting mean velocities and their corresponding uncertainties are tabulated in Table 1.

### 2.3 The pattern speed

Table 1 contains all the information that we need to calculate the pattern speed of the bar in NGC 936. As equation (1) shows, the pattern speed can be calculated independently from each of the values of $\langle x \rangle$ and $\langle \overline{v}_{\mathrm{los}}\rangle$ for the offset pointings relative to the values obtained for the nuclear spectrum. It is, however, more instructive to plot $\langle \overline{v}_{\mathrm{los}}\rangle$ versus $\langle x \rangle$ for all five pointings – if the analysis is correct, then equation (1) requires that these points lie on a straight line whose slope is $\Omega_p \sin i$. Figure 4 shows that the points do, indeed, approximately follow a straight line. A linear regression on these data gives a best-fit slope of

$$\Omega_p \sin i = 3.1 \pm 0.75 \ \mathrm{km\,s^{-1}\,arcsec^{-1}}. \quad (8)$$

The quality of the straight-line fit in Fig. 4, and the reasonable value of $\chi^2$ (1.6 with 3 degrees of freedom) imply that the observations from different slit positions are all consistent with the single bar pattern speed of equation (8). Further, we can use the $\chi^2$ statistic to conclude that the error analysis returns a realistic measure of the uncertainty in each of the estimates of $\langle \overline{v}_{\mathrm{los}}\rangle$.

It is also heartening that essentially identical slopes are derived by just considering the data from the pair of slit positions 14 arcseconds from the major axis or the pair 28 arcseconds from the major axis: TW pointed out that by using data in such pairs, systematic errors due to any uncertainty in the systemic velocity and centre of the galaxy vanish. Such pairing also eliminates any systematic error in the mean velocity which might arise from imperfect charge transfer in the CCD and other factors which might distort the galaxy's broadening function. Note, however, that the symmetry of the major axis broadening function (see Fig. 3) means that such skewing effects must, in any case, be small. A further possible source of systematic error can arise from "template mismatch" – the spectrum of the template star used in the analysis of Section 2.2 might be significantly different from that of the galaxy (and the galaxy's spectrum might also vary with radius). However, we have repeated the kinematic analysis using stellar templates with a range of spectral types, and we find that the derived values of $\langle \overline{v}_{\mathrm{los}}\rangle$ are insensitive to differences between the spectra of the galaxy and the templates. Similarly, poor subtraction of the sky contribution from the spectra could produce systematic errors in the derived kinematics, but varying the region of the slit from which the sky spectrum was extracted makes no significant difference to the derived values of $\langle x \rangle$ and $\langle \overline{v}_{\mathrm{los}}\rangle$. Hence, the value that we derive for $\Omega_p$ is not compromised by these possible systematic effects.

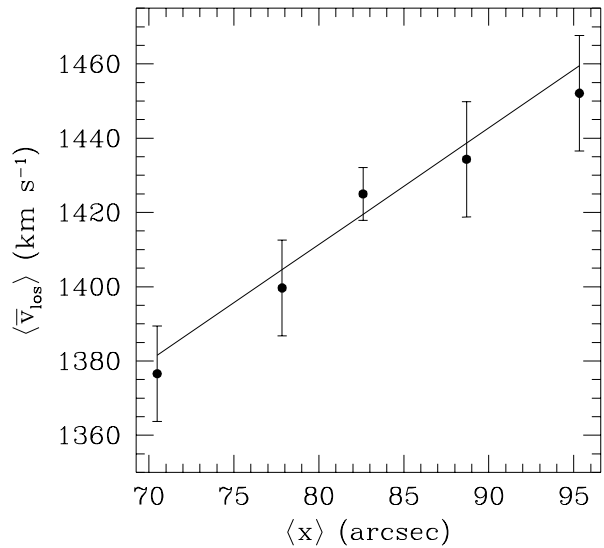

**Figure 4.** Mean line-of-sight velocity from the co-added spectrum as a function of the weighted mean position along the slit. The best-fit regression line weighted by the errors in $\langle \overline{v}_{\mathrm{los}}\rangle$ is indicated; the slope of this line gives the pattern speed.

## 3 DISCUSSION

The good fit to a straight line in Fig. 4 provides direct evidence for the validity of the assumptions in the TW method. It confirms that the continuity equation is satisfied for the stars in NGC 936, which is to be expected in this SB0 system since it contains no evidence for dust obscuration or on-going star formation. The analysis also shows that the bar seems to be a long-lived feature with a single well-defined pattern speed: if there were significant shearing effects in the bar which might arise from dynamical friction forces that vary with radius, then the bar would evolve rapidly by winding up, and Fig. 4 would no longer show a straight-line relation between $\langle x \rangle$ and $\langle \overline{v}_{\mathrm{los}}\rangle$ (Engström 1991).

The value that we obtain for $\Omega_p \sin i$ [equation (8)] is somewhat lower than Kent's (1987) estimate of $5.4 \pm 1.9 \ \mathrm{km\,s^{-1}\,arcsec^{-1}}$. However, the combined error on these two measurements means that the difference is not statistically significant. By using higher signal-to-noise ratio CCD spectra, and by coadding along the slit and analyzing the full galaxy broadening function, we have been able to reduce the uncertainty in the measurement considerably.

Using the disk inclination listed by Kormendy (1984) of $i = 41°$, the value for the bar pattern speed in the plane of NGC 936 is $\Omega_p = 4.8 \pm 1.1 \,\mathrm{km\,s^{-1}\,arcsec^{-1}}$. At an assumed distance of 16.6 Mpc (Kormendy 1984), this value corresponds to $60 \pm 14 \,\mathrm{km\,s^{-1}\,kpc^{-1}}$. Combining our pattern speed estimate with Kormendy's (1984) rotation curve for this galaxy, we can infer that co-rotation lies at a radius of $69 \pm 15$ arcsec. Kent & Glaudell (1989) found that the bar extends to a radius of $\sim 50$ arcsec (see also Fig. 1); we therefore find that the bar ends at a point close to, but within, its own co-rotation radius, in good agreement with the predictions of N-body simulations (e.g. Combes & Sanders 1981, Sellwood 1981). The result also fits very well with Kent & Glaudell's (1989) orbit calculations for the bar of NGC 936, since they found that they could best model the structure of this system if its pattern speed were $64 \,\mathrm{km\,s^{-1}\,kpc^{-1}}$. Any significantly higher pattern speed is essentially ruled out by their modelling, since it would result in a bar which extends to beyond its co-rotation radius: such bars are very difficult to build self-consistently since the principal orbit family outside the co-rotation radius is aligned perpendicular to the bar.

By applying new analysis techniques to high quality spectral data, we have shown that it is possible to derive stellar kinematic properties of galaxies with reasonable precision and well-defined uncertainties. It therefore now appears practical to use the Tremaine-Weinberg formula in order to obtain direct, accurate measurements of the pattern speeds of early-type barred galaxies. In the case of NGC 936, our findings are in good agreement with the theoretically-driven belief that galactic bars end just inside their co-rotation radii.

## ACKNOWLEDGMENTS

We thanks Marc Balcells and Marijn Franx for a thorough reading of the manuscript. The data presented in this paper were obtained using the Multiple Mirror Telescope, which is a joint facility of the Smithsonian Institute and the University of Arizona. Much of the analysis was performed using IRAF, which is distributed by NOAO, and some of the computation was carried out at the Southampton University Starlink node. MM is supported by a PPARC Advanced Fellowship. KK acknowledges travel support from the EU Human Capital and Mobility Programme.

## REFERENCES

Bender, R. 1990, A&A, 229, 441
Buta, R. 1986, ApJS, 61, 631
Combes, F., Gerin, M., 1985, A&A, 150, 327
Combes, F., Sanders, R.H., 1981, A&A, 96, 164
Contopolous, G., 1980, A&A, 81, 198
Engström, S., 1991, in Sundelius, B., ed., Dynamics of Disc Galaxies. Göteborgs University, Göteborg, p. 333
Franx, M., Illingworth, G.D., 1988, ApJ, 327, L55
Hernquist, L. & Weinberg, M.D., 1992, ApJ, 400, 80
Kent, S.M., 1987, AJ, 93, 1062
Kent, S.M., Glaudell, G. 1989, AJ, 98, 1588
Kormendy, J., 1984, ApJ, 286, 132
Kuijken, K., Merrifield, M.R., 1993, MNRAS, 264, 712
Sanders, R.H., Tubbs, A.D., 1980, ApJ, 235, 803
Sellwood, J.A., 1981, A&A, 99, 362
Sellwood, J.A., Wilkinson, A., 1993, Rep. Prog. Phys., 56, 173
Sempere, M.J., Garcia-Burillo, S., Combes, F., Knapen, J., 1994, A&A, in press.
Tonry, J.C., Davies, M., 1979, AJ, 84, 1511
Tremaine, S., Weinberg, M.D., 1984, ApJ, 282, L5
van der Marel, R.P., Franx, M., 1993, ApJ, 407, 525
Yahil, A., Vidal, N.V., 1977, ApJ, 214, 347